# Accessing ultrafast spin-transport dynamics in copper using broadband terahertz spectroscopy

Jiří Jechumtál[1*], Reza Rouzegar[2,3*], Oliver Gueckstock[2,3], Christian Denker[4], Wolfgang Hoppe[5], Quentin Remy[2,] Tom S. Seifert[2,3], Peter Kubaščík[1], Georg Woltersdorf[5], Piet W. Brouwer[2], Markus Münzenberg[4], Tobias Kampfrath[2,3], and Lukáš Nádvorník[1]

1. Faculty of Mathematics and Physics, Charles University, 121 16 Prague, Czech Republic
2. Department of Physics, Freie Universität Berlin, 14195 Berlin, Germany
3. Department of Physical Chemistry, Fritz Haber Institute of the Max Planck Society, 14195 Berlin, Germany
4. Institut für Physik, Universität Greifswald, 17489 Greifswald, Germany
5. Institut für Physik, Martin-Luther-Universität Halle, 06120 Halle, Germany

    * Authors contributed equally.

**Abstract:**

We study the spatiotemporal dynamics of ultrafast electron spin transport across nanometer-thick copper layers using ultrabroadband terahertz emission spectroscopy. Our analysis of temporal delays, broadening and attenuation of the spin-current pulse reveals ballistic-like propagation of the pulse peak, approaching the Fermi velocity, and diffusive features including a significant velocity dispersion. A comparison to the frequency-dependent Fick's law identifies the diffusion-dominated transport regime for distances $> 2\,\mathrm{nm}$. The findings lie the groundwork for designing future broadband spintronic devices.

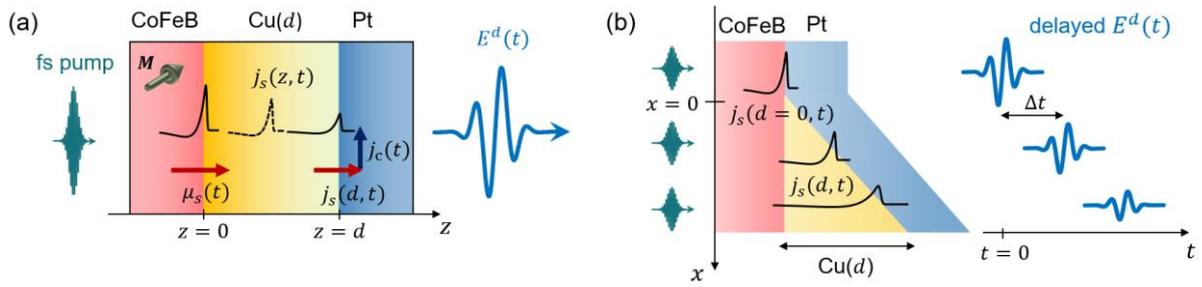

Fig. 1. THz spin-current generation, propagation and detection in a trilayer $F|Cu(d)|Pt$. (a) A fs laser pulse excites a ferromagnetic layer ($F = CoFeB$, in-plane magnetization $M$, green arrow) and injects a spin-current pulse $j_s(z = 0, t)$ (red arrow) into an intermediate layer $Cu$ with thickness $d$, where it undergoes attenuation and dispersion. Finally, $j_s$ is converted into a transverse charge current $j_c(t)$ in the Pt detection layer by the inverse spin Hall effect and radiates a THz pulse with electric field $E^d(t)$. (b) Sketch of the wedge design of the sample, showing the delayed, attenuated and broadened signals $j_s(d, t)$ and $E^d(t)$ for different $d$ as selected by the lateral position $x$ of the pump focus.

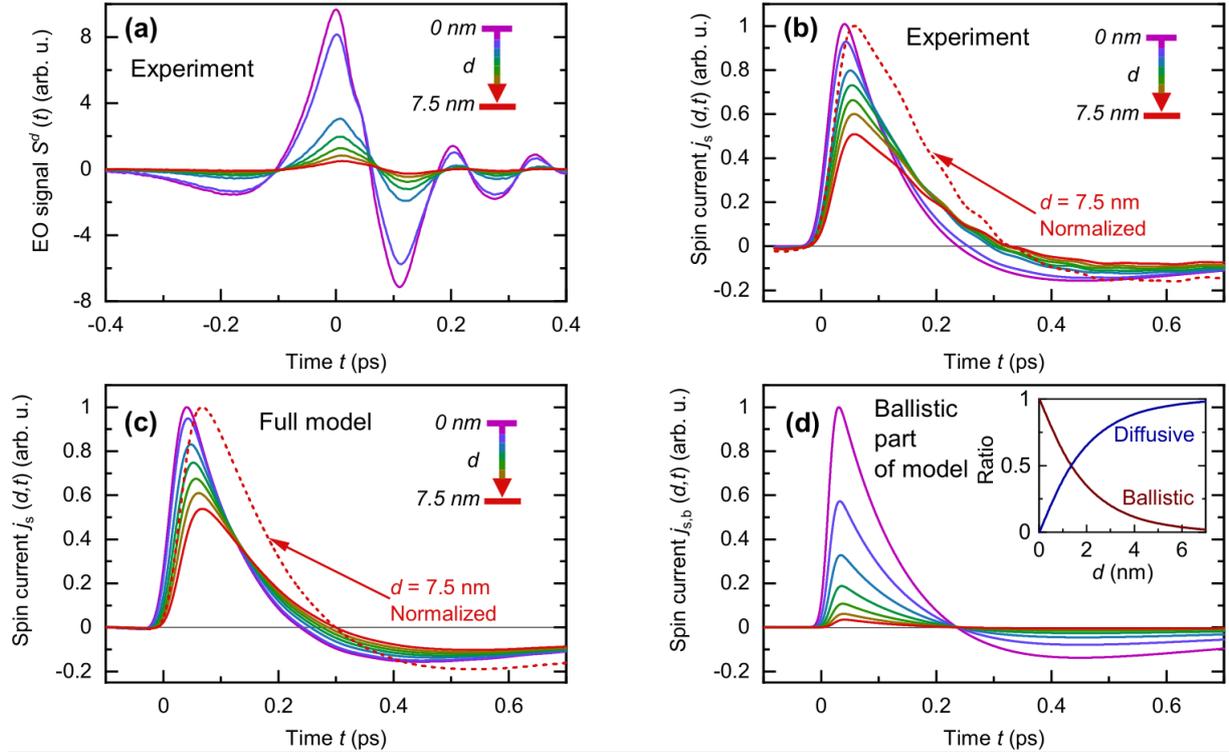

Fig. 2. THz spin currents $j_s(z, t)$ after traversing the Cu interlayer. (a) Measured THz emission signals $S^d(t)$ from $F|\text{Cu}(d)|N$ stacks for $d = 0, 0.6, 2.2, 3.4, 4.7, 5.9, 7.5$ nm (color-coded). (b) Extracted spin-current dynamics $j_s(z = d, t)$ according to Eqs. (1) and (2). The dashed black arrow indicates the delay $\Delta t$ of the current peak. (c) Spin currents $j_s(z = d, t)$ calculated using the spin-propagation model of Eqs. (1), (4), (5) with $v_F = 1.1$ nm/fs and $\tau = 4$ fs. (d) Calculated ballistic contribution to $j_s(z, t)$ by taking $\tau_{sf} = \tau$ in Eq. (2). Red dashed curves in (b), (c) are $j_s(z = d, t)$ for $d = 7.5$ nm with maximum normalized to unity. Inset: ratio of ballistic and diffusive components in $j_s(z, t)$ vs $d$.

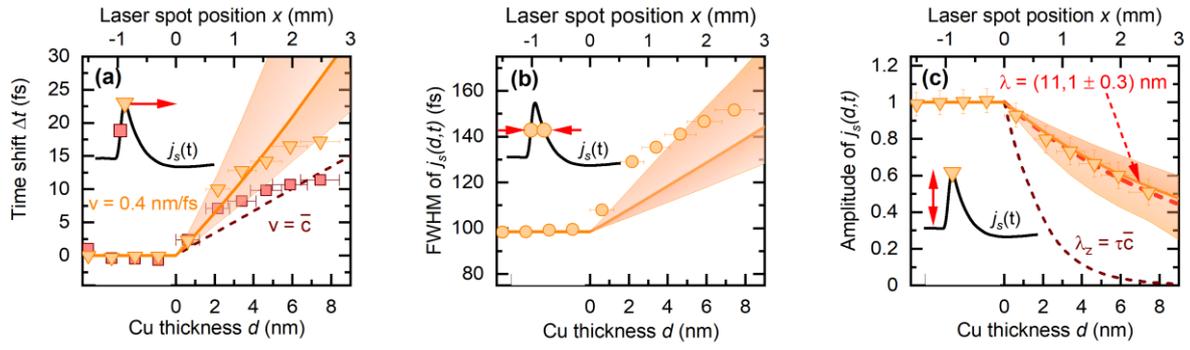

Fig. 3. Analysis of $j_s(d,t)$ from Fig. 2(b). (a) Delay $\Delta t$ of the peak (orange triangles) and leading edge (red squares) of the spin current $j_s(z=d,t)$ as a function of Cu thickness $d$. (b) Full width at half maximum (FWHM) (circles) and (c) peak amplitude of $j_s(z=d,t)$ vs $d$. In all panels, results of the model with $v_F = 1.1 \pm 0.2$ nm/fs and $\tau = 4 \pm 2$ fs are shown by orange solid curves (central values) and orange shades (uncertainties). The results of the ballistic contribution are shown by dark-red dashed curves.

Following the rapid development of terahertz (THz) and antiferromagnetic spintronics [1,2], THz spin currents (TSCs) are expected to play an essential role in future ultrafast spintronic devices [3]. For example, very recently, THz-pulse-driven TSCs were used to manipulate an antiferromagnetic memory bit on sub-picosecond time scales [4]. Another and complementary trigger of TSCs is optical excitation of thin-film multilayers by femtosecond laser pulses. This approach was successfully used for ultrafast spin-torque generation [5,6,7,8,9,10,11] or spintronic THz emission [12,13,14,15,16,17,18,19,20]. The latter concept has also found utility in THz investigation of formation [21,22,23] and dynamics [24,25,26,27,28] of ultrafast spin transport itself.

Following the theory works on TSCs [29,30,31,9], there is a rising number of experimental studies in the last years. For example, previous experiments inferred the temporal dynamics of a TSC after traversing distance $d$ from its impact on the magnetization of an adjacent layer [5,7,8] or by optical second-harmonic generation [5,32]. Other works also addressed the spatial evolution of TSCs, i.e., the amplitude reduction of emitted THz pulses with increasing $d$, and deduced the relaxation length of the underlying TSCs [27,33]. However, to reveal the complex propagation character of the ultrafast spin transport, direct experimental detection of the entire spatiotemporal evolution of the TSC dynamics, including its absolute temporal delay is required.

In this Letter, we investigate such spatiotemporal evolution of sub-picosecond spin-current pulses through a thin copper layer of thickness $d$ using time-domain THz emission spectroscopy with a high temporal resolution of 40 fs. By analyzing the THz signals, we directly infer the propagation speed of TSC pulses, their broadening and attenuation with $d$. We observe a ballistic-like propagation of the leading edge and peak of the TSC pulse with a speed approaching the Fermi velocity of Cu. The TSC pulse duration is found to increase by a factor >1.5 over a distance of $d = 8$ nm. Using a simple model based on a frequency-dependent Fick's law, we extract intrinsic spin-transport parameters, identify the dispersion of propagation velocities due to electron scattering as the source of the TSC pulse broadening and reveal diffusion as dominant spin-transport regime for $d > 2$ nm.

Our general approach to the generation, propagation and detection of a TSC is shown in Fig. 1. The sample is a $F|X|N$ trilayer [15] where $F = $ CoFeB is a ferromagnetic thin film, $N = $ Pt is a heavy metal with a large spin-Hall angle $\theta_{\mathrm{SH}}$, and $X = $ Cu($d$) is the interlayer with thickness $d$. First, a femtosecond laser pulse excites the $F|$Cu($d$)$|N$ sample and deposits the fraction $A_\mathrm{F}^d$ of the incident pump-pulse energy in $F$. It generates an ultrafast spin voltage $\mu_\mathrm{s}^d(t)$ in $F$ [21]. As a consequence, a sub-picosecond spin-current pulse $j_s(z=0,t)$ is launched from $F$ into the intermediate Cu layer [21], where $z$ is the out-of-plane coordinate (Fig. 1) and $t$ is time. The amplitude of $\mu_\mathrm{s}^d(t)$ is assumed to scale with $A_\mathrm{F}^d$ and the corresponding normalized dynamics $\mu_\mathrm{s}(t) = \mu_\mathrm{s}^d(t)/A_\mathrm{F}^d$ to be $d$-independent. Second, the spin current propagates inside the $X$ layer and undergoes attenuation and dispersion. We assume the propagation of the TSC proceeds in the linear-response regime and, thus, can be described inside $X$ by the convolution relation [34]

$$j_\mathrm{s}(z,t) = (\mu_\mathrm{s} * P)(t) = \int \mathrm{d}t'\, \mu_\mathrm{s}(t') P(z, t-t'). \tag{1}$$

Here, the response function $P(z,t)$ is the spin-current density that would be obtained for a $\delta(t)$-like spin voltage. Third, the current $j_s(z=d,t)$ arriving at the Pt detection layer is converted into a total transverse charge current $j_c(t) \propto \theta_{\mathrm{SH}} j_s(d,t)$ via the inverse spin Hall effect. We choose $N = $ Pt because its large inverse spin Hall effect dominates all other spin-to-charge-current conversion processes in the system [35,36,37]. Finally, the $j_c(t)$ emits a THz pulse with the electric field

$$E^d(t) = e Z^d A_\mathrm{F}^d j_c(t) \propto Z^d A_\mathrm{F}^d j_s(z=d,t), \tag{2}$$

which is detected. Here, $Z^d$ is the frequency-independent impedance of the sample. By measuring $E^d(t)$ for $d=0$, we obtain $\mu_\mathrm{s}(t)$ from Eqs. (2) and (1). By increasing $d$, one can approximately reconstruct $j_s(z=d,t)$ in the Cu spacer.

We summarize that our interpretation of the evolution of $j_s$ using Eqs. (1) and (2) relies on the following assumptions: (i) $j_s(z=0,t)$ originates solely from $\mu_\mathrm{s}^d(t)$ in $F$ [34]. (ii) Its amplitude scales with $A_\mathrm{F}^d$ [21,33,35]. (iii) The presence of the Cu layer does not change the dynamics of $\mu_\mathrm{s}(t)$ in $F$ [see Supplementary [38], Fig. S1]. (iv) $j_s(d,t)$ is fully absorbed and converted inside $N$ [33]. (v) The measured $j_c(t)$ is exclusively due to spin-charge conversion in $N = $ Pt, i. e., $\theta_\mathrm{SH} \neq 0$ only inside $N$ [24,35]. It follows that the values of $Z^d$ and $A_\mathrm{F}^d$ impact only the amplitude of $j_s(d,t)$, not its dynamics.

In the experiment, we measure an electro-optical THz signal that is related to $E^d(t)$ by the convolution relation

$$S^d(t) = (H * E^d)(t). \quad (3)$$

Here, $H(t)$ is the setup transfer function that can be determined experimentally [39,40]. Using Eq. (3), $E^d(t)$ is retrieved by the deconvolution procedure detailed in Supplementary [38], Note 1 [24,25].

Our $F|Cu|N$ stack has the layer structure Co$_{40}$Fe$_{40}$B$_{20}$(2 nm)|Cu($d$)|Pt(2 nm) and is grown by electron-beam evaporation on a double-side polished Al$_2$O$_3$ substrate [Fig. 1(a)]. On half the sample area, the Cu layer forms a wedge with a gradient of $\partial d/\partial x = 3.1$ nm/mm along the $x$ direction [Fig. 1(b)] (see Supplementary [38], Note 2 for details). This configuration allows us to conveniently select $d(x)$ in the range between 0 and 7.5 nm by positioning the pump laser beam at the appropriate position $x$. The other part of the sample lacks the wedge ($d = 0$) and forms a stripe of $F|N$ used for referencing.

To access relative amplitudes of $j_s(z,t)$ in Eq. (2), the sample is characterized for all $d$ in terms of the absorbed pump-pulse energy fraction $A_\mathrm{F}^d$ in $F$ and the total impedance $Z^d$ using THz transmission spectroscopy [both detailed in Supplementary [38], Note 3 and Fig. S2]. The measured $Z^d$ are found to be almost frequency-independent up to 7 THz for all $d$ [Fig. S2(b) and (c)] and decrease steeply with $d$ [Fig. S2(d)]. For $d = 11$ nm, the stack conductance $G$ is dominated by the Cu layer (Supplementary [38], Note 3) and, thus, the Drude model [40,41] $G(\omega) \propto 1/(1 - i\omega\tau)$ can be used to estimate the electron scattering time $\tau$ in Cu. Indeed, it provides a good fit to the data for $1 < \tau < 10$ fs [Fig. S2(f)]. This $\tau$ is much smaller than the several tens of fs typical for epitaxial Cu layers [42,43] and assigned to the polycrystalline nature of our evaporation-deposited wedge.

In the THz emission experiments, the sample is excited by a train of ultrashort laser pulses (wavelength 790 nm, duration 10 fs, repetition rate 80 MHz, energy per pulse 2 nJ) from a Ti:Sapphire laser oscillator. The pump beam is focused to a spot with a full width at half maximum (FWHM) of the intensity of ~30 μm on the sample. Its lateral position $x$ sets $d$ with a precision of ~0.1 nm [Fig. 1(b)]. The magnetization $\boldsymbol{M}$ of the $F$ layer is controlled by an external magnetic field of ~10 mT. The emitted THz electric field $E(t)$ propagates in a nitrogen atmosphere and is detected as an electrooptical (EO) signal $S(t)$ via EO sampling [44,45] in a 250 μm thick GaP(110) crystal by using linearly polarized probe pulses (0.6 nJ) split from the pump beam. Even though the temporal resolution of the subsequent analysis is ~40 fs, a continuously scanning delay line together with a high signal-to-noise ratio of our setup allows us to resolve the minimal increment of time $t$ and, thus, also the temporal delays $\Delta t$ of $S(t)$, as fine as 1.6 fs.

Typical THz-emission waveforms $S^d(t)$ from the CoFeB|Cu($d$)|Pt samples are shown in Fig. 2(a). As $d$ increases from 0 to 7.5 nm, the overall signal amplitude decreases by roughly a factor of 20. The noticeable reduction between the second and third curve originates from the non-equidistant choice of $d$ and the steep nonlinear decrease of $Z^d$, while the oscillatory features at $t > 0.2$ ps arise from the EO detection process (Supplementary [38], Note 1). Interestingly, the absolute maximum and minimum of $S^d(t)$ undergo a gradual temporal shift $\Delta t$ (dashed arrows), where $\Delta t$ appears to be larger for the minimum. We experimentally rule out possible trivial sources of time delays in $S^d(t)$ such as the Gouy phase shift, a variation of the substrate thickness and long-term temporal drifts of the laser (see Supplementary [38], Note 4). We note that all waveforms are antisymmetrized with respect to $\boldsymbol{M}$, i.e., $S^d(t) = [S^d(+\boldsymbol{M},t) - S^d(-\boldsymbol{M},t)]/2$, which captures only relevant effects odd in $\boldsymbol{M}$ (see Supplementary [38], Figs. S3 and Note 5 for further confirmation of the spintronic origin of signal).

To extract the TSC density $j_s(z = d, t)$ directly behind the Cu layer, we apply Eqs. (2) and (3) to $S^d(t)$ normalized by $Z^d A_\mathrm{F}^d$ [see the mono-exponential amplitude decrease in Supplementary [38], Fig. S4 as compared to Fig. 2(a)] and use assumptions (i-v) (Supplementary [38], Note 1). Figure 2(b) shows the resulting $j_s(z = d, t)$ for various values of $z = d$ and, thus, provides the approximate spatial evolution of the ultrafast dynamics of the TSCs. The rise time of $j_s(d = 0, t)$ indicates that the time resolution of the extracted TSCs is ≈40 fs. Both the gradual attenuation and the rising temporal shift $\Delta t$ of the THz signals $S^d(t)$ vs $d$ are preserved in the TSCs [arrow in Fig. 2(b)]. Importantly, $j_s(d,t)$ undergoes a notable broadening that is clearly visible for $j_s(d = 0, t)$ vs the normalized $j_s(d = 7.5\,\mathrm{nm}, t)$ (red dashed line) without any further analysis. This behavior indicates that the TSC undergoes a significant dispersion.

To quantify these qualitative observations, we look at the details of the measured spin current by extracting temporal delay $\Delta t$, the full width at half maximum (FWHM), and amplitude of the spin current pulses. First, we extract $\Delta t$ relative to $j_s(d = 0, t)$ for the leading edge (at half maximum) and peak of the spin current for all measured $d$. The resulting $\Delta t$ vs $d$ [Fig. 3(a)] shows a monotonic nonlinear increase. The propagation velocity of the leading edge and peak of $j_s(d, t)$ can be obtained from fitting the mean slope of $\Delta t(d)$ [Supplementary [38], Fig. S5(a)], yielding propagation speeds of, respectively, $(0.6 \pm 0.1)$ nm/fs and $\approx 0.4$ nm/fs for the whole data range and, correspondingly, $\approx 1.2$ nm/fs and $\approx 0.7$ nm/fs for $d > 2$ nm [Fig. S5(b)]. The latter values are sizable and almost reach the Fermi velocity ($\sim 1.1$ nm/fs [43]) of electrons in Cu. Moreover, the pulse leading edge seems to propagate faster than the subsequent pulse peak. This behavior implies a broadening of the leading edge and, possibly, the whole TSC pulse $j_s(d, t)$ as $d$ increases.

Second, the FWHM of $j_s(d, t)$ vs $d$ is shown in Fig. 3(b). Indeed, we find a significant pulse broadening from 100 fs at $d = 0$ by a factor of 1.5 at $d = 7.5$ nm of Cu. Third, the amplitude of the peak TSC decreases exponentially with $d$ with a relaxation length of $\lambda_{\mathrm{rel}} = (11.1 \pm 0.3)$ nm [Fig. 3(c)]. We note that the abrupt amplitude reduction visible in $S^d(t)$ around $d = 2$ nm [Fig. 2(a)] disappears in amplitudes of $j_s(d, t)$ due to normalization by $Z^d A_F^d$. Therefore, the proper $\lambda_{\mathrm{rel}}$ cannot be retrieved directly from $S^d(t)$ and underscores the importance of the reconstruction of $j_s(d, t)$ [see Supplementary [38], Fig. S4(b)].

To better understand the observed TSC-pulse dynamics [Figs. 2(b) and 3(a-c)], in particular its edge and peak delay and temporal broadening, we make use of an analytical model of ultrafast spin transport in Cu [29,30]. It relies on two macroscopic relationships that can be derived from the Boltzmann transport equation. (i) In a generalized version of Fick's law, $j_s(z, \omega) = -D_s(\omega)\partial_z \mu_s$, the diffusion coefficient is proportional to the Cu conductance. Thus, it has the same Drude-type frequency dependence, $D_s(\omega) \propto G(\omega) \propto 1/(1 - i\omega\tau)$, with $\tau$ estimated in Supplementary Fig. S2(f). (ii) In the time-domain continuity equation, $\partial_z j_s \propto -\partial_t \mu_s - 2\mu_s/\tau_{\mathrm{sf}}$, the second term is due to spin relaxation in Cu [46,47,48,49].

By combining (i) and (ii) (see Supplementary [38], Note 6), we find that the response function $P$ [see Eq. (1)] can be written as the Fourier integral

$$P(z, t) = \int \mathrm{d}\omega\, \mathrm{e}^{ik(\omega)z - i\omega t}. \tag{4}$$

Here, for each frequency $\omega/2\pi$, the associated complex-valued wavevector $k$ is given by the dispersion relation $\bar{c}^2 k^2 = \omega^2 + i\omega(\tau^{-1} + 2\tau_{sf}^{-1}) - 2(\tau\tau_{\mathrm{sf}})^{-1}$, where $\bar{c} = v_F/\sqrt{3}$ is the mean electron band velocity projected on the $z$ direction. By considering that $\tau_{\mathrm{sf}} \gg \tau$ [46,47,48,49], this relation simplifies to

$$\frac{\bar{c}^2 k^2}{\omega^2} = 1 - \frac{1}{i\omega\tau}. \tag{5}$$

Eqs. (4) and (5) allow us to interpret TSC-pulse propagation through Cu as signal transmission [29]. The frequency-dependent group velocity $\partial_k \omega$ and attenuation $\mathrm{Im}\, k(\omega)$ follow from the dispersion relation [Eq. (5)]. Note that the model captures both ballistic and diffusive transport, which prevail, respectively, for angular frequencies $\omega$ much larger and smaller than the rate $\tau^{-1}$ of electron momentum scattering. For example, for $\omega \gg \tau^{-1}$, we can neglect the second term on the right-hand side of Eq. (5), and the TSC-pulse group velocity approaches the mean electron band velocity $\bar{c}$.

We use Eqs. (1), (4) and (5), the experimentally given $\mu_s(t)$, and the known $v_F = 1.1$ nm/fs [43] and the mean $\tau = 4$ fs [Supplementary [38], Fig. S2(f)] to calculate the resulting TSC dynamics. The calculated $j_s(z = d, t)$ [Fig. 2(c)] agree well with the measured $j_s(d, t)$ [Fig. 2(b)]. From the modeled dynamics, we extract the peak delay $\Delta t$, the pulse FWHM and peak amplitude of $j_s(z = d, t)$ as a function of $d$ and plot them as orange lines in Fig. 3(a)-(c). The orange shaded areas correspond to a small variation of $v_F$ by $\pm\, 0.2$ nm/fs and $\tau$ by $\pm\, 2$ fs. We find reasonably good agreement of model and experiment, showing that the transport features can be explained by a combination of ballistic and diffusive components. However, we find that the model underestimates the TSC pulse broadening and has a slight mismatch with $\Delta t$.

To obtain more insight into the role of electron scattering, we extract the ballistic component $j_{s,b}(z,t)$ of the calculated $j_s(z,t)$ by considering a special condition of $\tau_{sf} = \tau = 4$ fs and employ it in the exact dispersion relation $k(\omega)$. This choice spin-depolarizes all electrons that have experienced a scattering event and, thus, does not make them available for diffusive spin transport. The resulting ballistic component $j_{s,b}(d,t)$ [Fig. 2(d)] decays considerably faster with increasing $d$ than the diffusive component $j_s(z,t) - j_{s,b}(z,t)$, as also documented by the ratio of the 2 contributions [inset in Fig. 2(d)]. For comparison, the parameters $\Delta t$ and attenuation of the maximum of the ballistic component are shown in Fig. 3(a) and (c) as dark-red dashed curves. We see that the attenuation and broadening of the measured TSC pulses cannot be explained by scattering-free spin transport and requires a non-ballistic component. Indeed, the observed amplitude relaxation length of $\lambda_{rel} \sim 11$ nm is about 5 times larger than the mean free-path of $\lambda_z = \bar{c}\tau \approx 2$ nm along $z$. Fig. 2(d) and 3(c) also highlight that the diffusion (scattering-based) transport modes dominate the propagation for $d > 2$ nm.

Interestingly, the propagation speed of the TSC pulse front or its peak still reaches values close to $v_F$ and almost matches the wave-front velocity $\bar{c}$ expected from the model. This observation indicates that the leading parts of the TSC pulse are formed by electrons that experience only a few collisions, allowing for ultrafast (ballistic-like) spin propagation over length scales of more than 10 nm, not strictly limited by $\lambda_z$. To directly observe modes propagating at the speed $v_F$, it would be necessary to fulfill $\omega > 1/\tau$ by using materials with significantly larger $\tau$ by increasing the bandwidth of our experiment and analysis.

The role of the particular spin-transport regime or different layer quality, i. e., different $\tau$, is manifested in the literature by an interestingly large variation of $\lambda_{rel}$ for the spin transport in Cu: ranging from 4 nm [33], through 11 nm [50] and 50 nm [27], up to 120 nm [8] in the ultrafast transport experiments, and even larger values of the order of hundreds of nm are reported at GHz and lower frequencies [27,47,51]. It can be understood within the theory in Refs. 29 and 30 which shows that the diffusive spin transport, observed in the limit of $\omega \sim 0$, becomes ballistic when $\omega\tau \gg 1$. The ballistic spin transport by conduction electrons is limited by the mean free path at which, statistically, the first scattering event occurs. On the other hand, the diffusive spin transport allows for a further random-walk-like transport which extends to the usually significantly larger spin diffusion length [52]. While the low frequency spin transport is of the diffusive nature, the ultrafast/THz experiments study broadband spin-current pulses and, thus, the measured spin-current decay length reflects an average $\lambda$, ranging between the mean free path and the spin diffusion length.

The source of the visible underestimate of the TSC pulse broadening by the used model [Fig. 3(b)] could lie in disregarding a possible initial velocity distribution at $t = 0$. Indeed, the varying $z$-component of the initial velocity $v_z = v_F \cos(\theta)$ of electrons moving at angle $\theta$ from the out-of-plane $z$-axis might also induce an effective velocity distribution, not included in the model, and it can lead to an additional broadening of $j_s(z = d, t)$ [53,5]. However, if we apply the ballistic-only model with a homogeneous initial distribution of $\theta$ described in Methods in Ref. 53, it would induce a broadening by only a factor of ~1.2 over 8 nm of Cu, i.e., a significantly smaller value than what was observed in the measured dynamics. In order to include a more realistic initial $v_z$-distribution in the model, one would need to analyze the orbital symmetry matching between CoFeB and Cu. Because of these limitations, the model does not capture the slightly non-linear trend in Fig. 3(a) and (b) for $d < 3$ nm and, thus, invites for detailed studies of $\mu_s(t)$ at extreme interlayer thicknesses.

In summary, we employed time-domain THz emission spectroscopy to directly measure the spatial and temporal evolution of ultrafast spin currents triggered by optical excitation of metallic thin films. The observed temporal delays, significant broadening and attenuation of TSCs for varying Cu spacer thickness indicate diffusion-dominated spin transport and related dispersion of TSCs. A simple model based on the dynamic diffusion equation explains very well our data by assuming realistic values $v_F = 1.1$ nm/fs and $\tau = 4$ fs. If confirms the dominant role of electron-scattering in TSCs for thicknesses $d > 2$ nm. Notably, the analysis of the TSC pulse font revealed that the spin-current speed approaches the Fermi velocity. Our methodology facilitates practical implementation of spin currents in ultrafast spintronic devices. For spintronic emitters [13,14,15,16], we anticipate that Cu intermediate layers can be used to tune the spin current profile and consequently the performance of the spintronic THz emission.

**Acknowledgments**

The authors thank Afnan Alostaz for help with measurements. The authors acknowledge funding by the Czech Science Foundation through projects GA CR (Grant No. 21–28876J), the Grant Agency of the Charles University (SVV–2024–260720), the Deutsche Forschungsgemeinschaft (DFG, German Research Foundation) through the Collaborative Research Center SFB TRR 227 "Ultrafast spin dynamics" (project ID 328545488, projects A05, B02, and B03), and the priority program SPP2314 INTEREST (project ITISA; Project No. KA 3305/5-1), and the European Research Council (ERC) through the H2020 CoG project TERAMAG (Grant No. 681917). R. R. acknowledges support from the International Max Planck Research School (IMPRS) for Elementary Processes in Physical Chemistry. J. J. acknowledges the support of the Grant Agency of Charles University (Grant No. 120324). P. K. acknowledges the support of the Grant Agency of Charles University (Grant No. 166123).

# Accessing ultrafast spin-transport dynamics in copper using broadband terahertz spectroscopy

Jiří Jechumtál[1*], Reza Rouzegar[2,3*], Oliver Gueckstock[2,3], Christian Denker[4], Wolfgang Hoppe[5], Quentin Remy[2], Tom S. Seifert[2,3], Peter Kubaščík[1], Georg Woltersdorf[5], Piet W. Brouwer[2], Markus Münzenberg[4], Tobias Kampfrath[2,3], and Lukáš Nádvorník[1]

1. Faculty of Mathematics and Physics, Charles University, 121 16 Prague, Czech Republic
2. Department of Physics, Freie Universität Berlin, 14195 Berlin, Germany
3. Department of Physical Chemistry, Fritz Haber Institute of the Max Planck Society, 14195 Berlin, Germany
4. Institut für Physik, Universität Greifswald, 17489 Greifswald, Germany
5. Institut für Physik, Martin-Luther-Universität Halle, 06120 Halle, Germany

* Authors contributed equally.

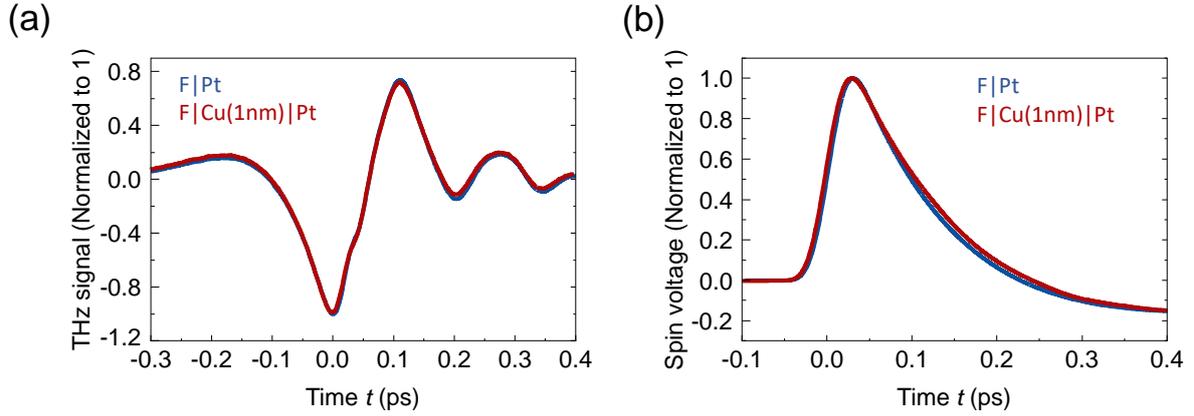

**Fig. S1. Spin voltage dynamics.** (a) The THz emission signals $S^d(t)$ with minimum normalized to $-1$ from $F|N$ and $F|\text{Cu}(d=1\,nm)|N$ stacks with $F = \text{CoFeB}(2\text{nm})$ and $N = \text{Pt}(2\text{nm})$. Both dynamics are almost identical. (b) Extracted spin currents $j_s(z=d,t)$ from THz signals shown in panel (a), showing an analogous overlap. Similarity in $S^d(t)$ and $j_s(d,t)$ indicates that Cu interlayer does not significantly change the spin voltage dynamics $\mu_s(t)$ in $F$. The small differences in $j_s(d,t)$ are mostly due to the spin transport in Cu.

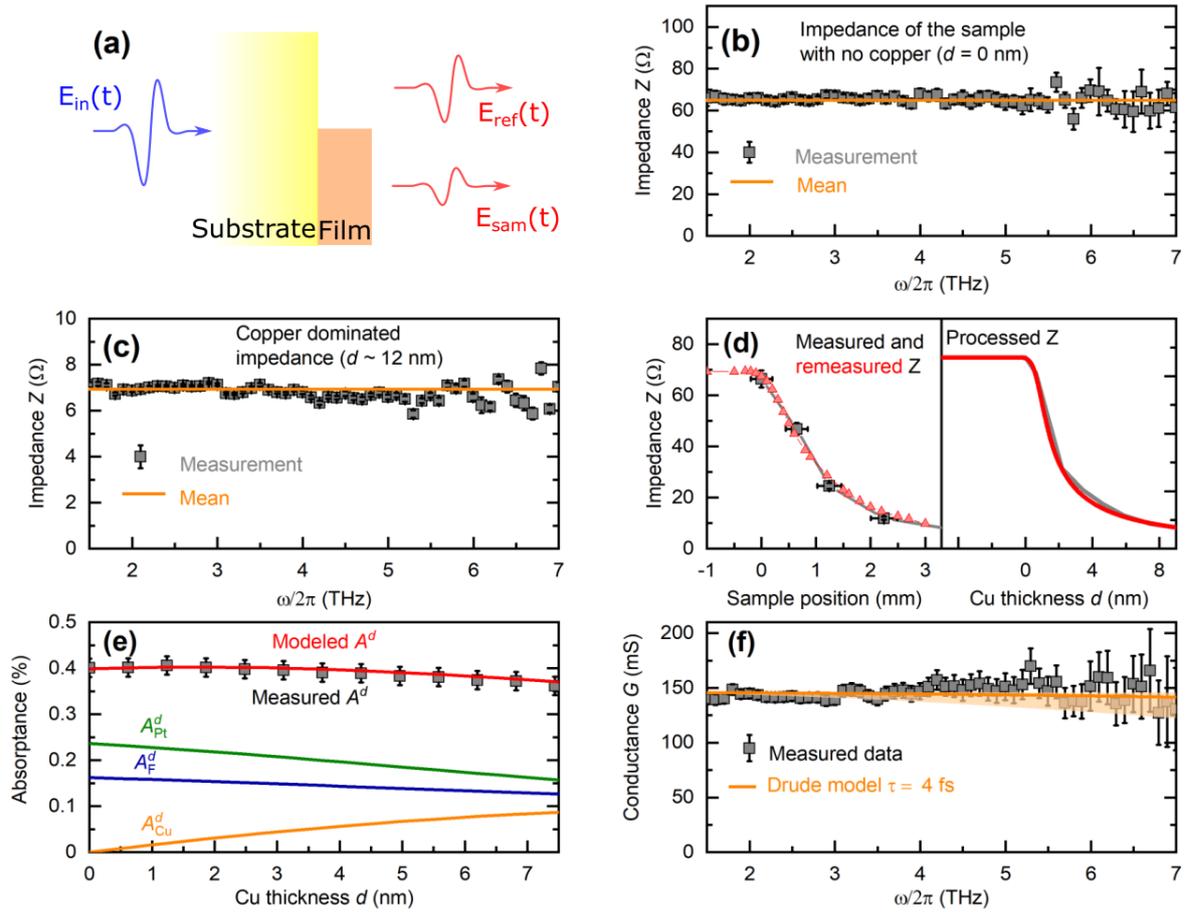

**Fig. S2. THz transmission spectroscopy.** (a) Schematic of THz transmission spectroscopy. The analysis is described in Note 3. (b) Measured frequency dependent THz impedance $Z(\omega)$ for the $F|Cu(d=0)|N$ and (c) $F|Cu(d=12\,nm)|N$ stacks with $F = \text{CoFeB(2nm)}$ and $N = \text{Pt(2nm)}$. $Z(\omega)$ is approximately constant in $\omega$ as shown by fits by a constant (orange horizontal lines), giving a mean $Z$ for each Cu thickness $d$. (d) Left: The extracted mean $Z(d)$ dependence for our full set of samples $F|Cu(d)|N$ from two independent experiments (grey and red points), showing an excellent reproducibility. Right: The measured $Z(d)$ processed by deconvolution of the transmission THz spot size (FWHM of 1.8 mm) and convolution by the emission THz spot size (0.3 mm), used in the normalization of emission data. (e) Measured (grey points) and calculated (red curve) optical absorptance $A^d$ of the whole $F|Cu(d)|N$ stack series. The calculated $A^d$ is the sum of the absorptance of Pt ($A_{Pt}^d$, green curve), CoFeB ($A_F^d$, blue curve) and Cu ($A_{Cu}^d$, orange curve). (f) Real part of the THz conductance $G(\omega)$ for $d = 11\,nm$, obtained by the procedure described in Note 3. It is compared to the fit by a Drude model for $\tau = 4$ fs (orange curve) and $1\,\text{fs} < \tau < 10\,\text{fs}$ (orange shaded area).

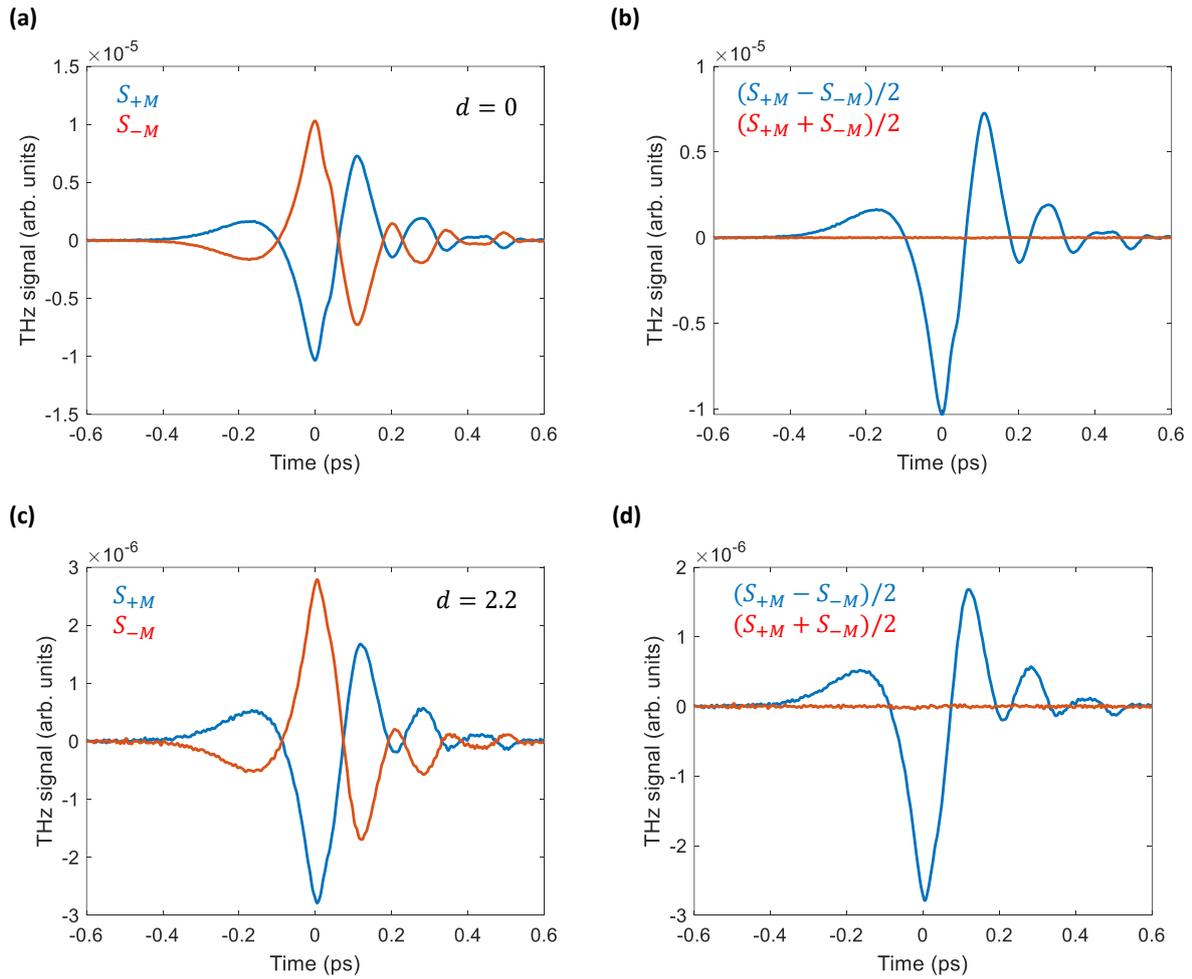

**Fig. S3. Magnetic and non-magnetic components of raw data.** (a) The raw EO signals measured for $d = 0$ and positive $(+M)$ and negative $(-M)$ polarity of magnetization. (b) The corresponding "magnetic" difference signal $S^- = (S_{+M} - S_{-M})/2$, odd in $M$, and the "non-magnetic" sum signal $S^+ = (S_{+M} + S_{-M})/2$, even in $M$. Note that $S^- \approx 200\, S^+$. (c,d) Same as (a,b) but for $d = 2.2\ nm$.

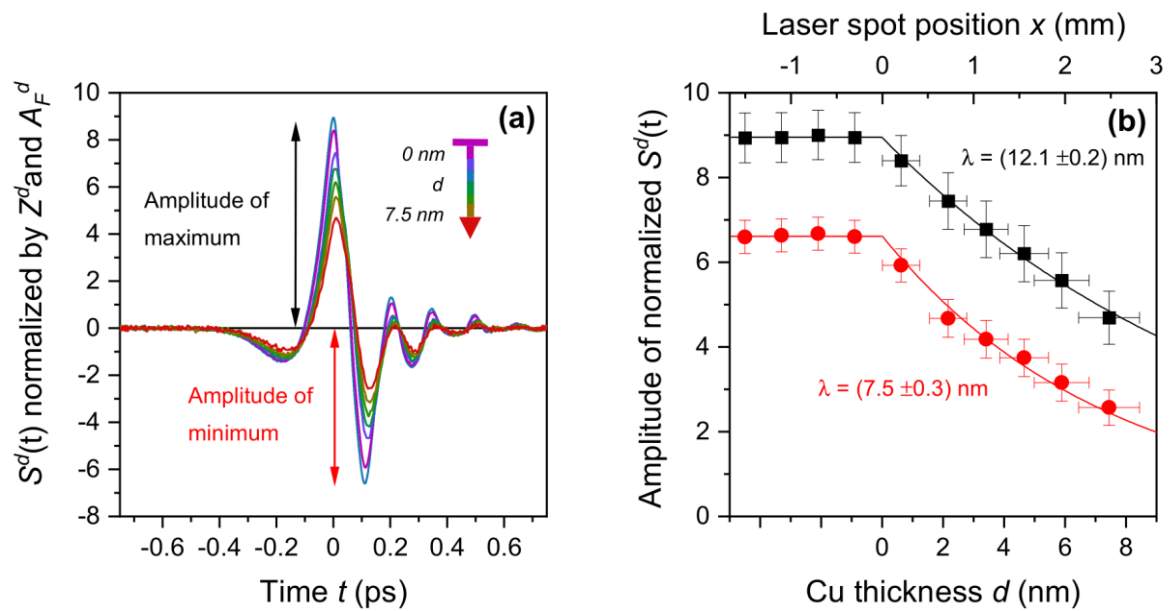

**Fig. S4. Normalization and relaxation lengths of THz EO signals.** (a) EO signals $S^d(t)$ normalized by the impedance $Z^d$ and the absorptance $A_F^d$, plotted for different Cu thicknesses $d$, showing the monotonous reduction of the amplitudes of maxima (black arrow) and minima (red arrow). (b) Corresponding amplitudes as a function of $d$ with simple exponential fits by $\exp(-d/\lambda)$. The parameters $\lambda$ differ considerably for different points of extraction $t$ and from the value obtained directly from the amplitude of the spin current $j_s(d,t)$, yielding $\lambda = (11.1 \pm 0.3)$ nm. This underscores the importance of the deconvolution of $S^d(t)$ into $j_s(d,t)$ before analyzing the amplitude reduction.

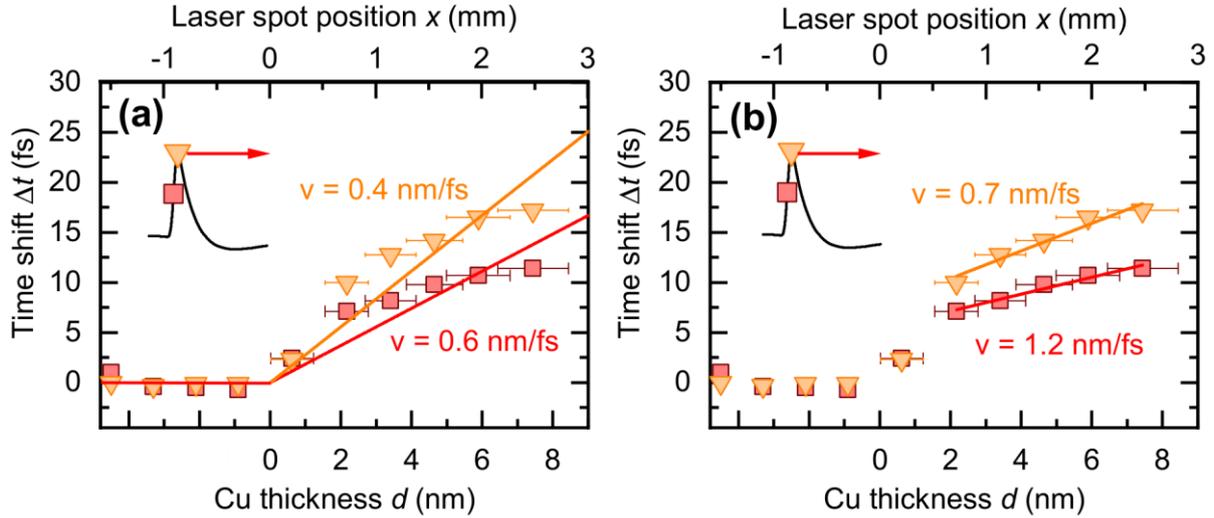

**Fig. S5. Fits of the displacement of the spin-current pulse.** (a, b) Delay $\Delta t$ of the peak (orange triangles) and leading edge (red squares) of the spin current $j_s(z=d,t)$ as a function of Cu thickness $d$. The fits are linear functions, estimating the slope of $\Delta t(d)$ in the whole range of $d$ (left panel) and for $d > 2$ nm (right panel), yielding the indicated spin-current propagation speeds.

## Supplementary Notes

### Note 1. Spin current extraction and deconvolution

**THz signal.** In our setup, we detect any transient electric field $E^d(t)$ for a particular Cu layer thickness $d$ by electro-optic (EO) sampling [1,2], where a probe pulse (0.6 nJ, 10 fs) co-propagates with the terahertz pulse through an electro-optic crystal. The ellipticity $S^d(t)$ accumulated by the sampling pulse is measured as a function of the time delay $t$ between the terahertz and sampling pulse by means of a polarization-sensitive optical bridge, which consists of a quarter-wave plate, a polarizing beam splitter and two balanced photodiodes. For the electro-optic crystal, we use GaP(110) (thickness of 250 µm). All experiments are performed at room temperature in a dry air atmosphere.

The EO signal $S^d(t)$ is related to the electric field $E^d(t)$ through the convolution relation

$$S^d(t) = (H * E^d)(t) \tag{S1}$$

where $H(t)$ is the transfer function which accounts for the THz pulse propagation to the detector and the electro-optic sampling process [3,4]. The transfer function $H(t)$ can be determined using a well-understood reference emitter, GaP(110), with a thickness of 50 µm [4]. One can find $E^d(t)$ by numerically solving Eq. (S1) while the convolution is time-discretized and recast in the form of a matrix equation, resulting in an estimated time resolution of 40 fs.

In the frequency domain, the convolution in Eq. (S1) becomes a multiplication, $S^d(\omega) = H(\omega)E^d(\omega)$, and $H(\omega)$ has the meaning of a filter function. If the bandwidth of $H(\omega)$ is narrower than that of $E^d(\omega)$, this filtering process leads to bandwidth reduction in $S^d(\omega)$ and subsequent formation of oscillatory features in the time-domain signal $S^d(t)$ [see Fig. 2(a)].

**THz spin current.** The measured electric field $E^d(t)$ for the sample $F|Cu(d)|N$ is related to the spin current $j_s(z = d, t)$ by the relation [5,6]

$$j_s(d, t) = \frac{E^d(t)}{Z^d A_F^d \theta_{\text{SH}} \lambda_{\text{Pt}}} \tag{S2}$$

where $Z^d$ is the frequency-independent impedance of the sample (see Note 3), $A_F^d$ I the fraction of pump absorbed in the $F$ layer, $\theta_{\text{SH}}$ is the inverse spin Hall angle of $N$, and $\lambda_{\text{Pt}}$ is the spin-current relaxation length in $N$.

### Note 2. Preparation of copper wedge interlayer

**Moving mask deposition.** The wedge has been produced by the electron-beam evaporation of Cu where the sample was gradually shadowed by a moving mask. Slow and steady linear motion of the mask during the evaporation process created a homogeneous wedge of a high quality and lateral homogeneity as reported in previous works [7–10] and allowed us to calculate the gradient of the wedge $\partial d/\partial x = 3.1$ nm/mm from the blade speed and evaporation rate. Here, $d$ is the Cu thickness and $x$ the lateral inplane coordinate [see Fig. 1(b)].

**Impact of the wedge angle on direction of $j_s(t)$.** The direction of the triggered spin current $j_s(t)$ is always perpendicular to the $F|Cu$ interface, so, strictly speaking, the angle of incidence of $j_s(t)$ on Pt layer is not normal. However, the formed wedge angle is extremely small. Using the known gradient $\partial d/\partial x$, we obtain the elevation angle $\alpha \approx 1.8 \times 10^{-4}$ degree. The incident direction of $j_s(t)$ to the platinum layer is, therefore, extremely close to normal and the travelled path equals to $d$ with a very high precision: $d\cos(\alpha) \approx d(1 - \alpha^2/2) \approx d$ where the second approximation neglects the extremely small terms smaller than $10^{-10}$ nm. The incident direction can be always considered to be normal even in case of small inhomogeneities of $\alpha$.

## Note 3. THz and optical transmission spectroscopy

**THz spectroscopy.** We employed the THz transmission spectroscopy to measure the impedance $Z_{\text{sam}}(\omega)$ and electrical conductivity $\sigma(\omega)$ of the thin film of our stacks.

Fig. S1(a) shows the design of the transmission arrangement. The THz pulse transmitted through the bare substrate, denoted $E_{\text{ref}}$, serves as a reference waveform while the pulse transmitted through the substrate and the thin film (our metallic stack), denoted $E_{\text{sam}}$, is the signal. Using the thin-film approximation [11], the impedance of the thin film is simply given by

$$Z_{\text{sam}}(\omega) = \frac{Z_0}{n_1(\omega) + n_2(\omega)} \frac{E_{\text{sam}}(\omega)}{E_{\text{ref}}(\omega)}, \tag{S3}$$

where $Z_0$ is the vacuum impedance, $n_1$ and $n_2$ are refractive indices of the substrate and the air. Note that $E_{\text{sam}}(\omega)/E_{\text{ref}}(\omega) = S_{\text{sam}}(\omega)/S_{\text{ref}}(\omega)$ where $S_{\text{sam}}$ and $S_{\text{ref}}$ are the measured electro-optic THz signals [see Eq. (S1)] [11].

Fig. S1(b) and (c) show the measured impedance of the sample F|Cu($d = 0$)|N and F|Cu($d = 12\ nm$)|N stack with $F = $ CoFeB(2nm) and $N = $ Pt(2nm). The film impedances are almost frequency-independent and $Z_{\text{sam}} \approx$ 6.5 Ω and 64 Ω, respectively, for the frequency range of 1-7 THz.

The film conductance $G(\omega)$ is obtained through the relation [11]

$$Z_{\text{sam}}(\omega) = \frac{Z_0}{n_1(\omega) + n_2(\omega) + Z_0 G(\omega)} \tag{S4}$$

and is shown for sample F|Cu($d = 11\ nm$)|N in Fig. S1(f), yielding the mean value $G_{\text{FCuN}} \approx 148$ mS. For this $d$, the conductance is dominated by the Cu conductance as the conductance of F|Cu($d = 0$)|N, obtained from Fig.S1(b) using Eq. (S4), yields the mean value of only $G_{\text{FN}} = 4.2$ mS. Therefore, the electron scattering time of Cu $\tau$ can be estimated by comparing $G(\omega)$ in Fig. S1(f) with the Drude model [11]. We show the model for $\tau = 4$ fs and for an interval of $1 < \tau < 10$ fs in Fig. S1(f) as orange curve and shades, respectively.

**Optical transmission.** The total stack absorptance $A^d$ [grey data points in Fig. S1 (e)] is experimentally determined by measuring the fraction of the reflected $R^d$, and transmitted $T^d$ pump power, resulting in $A^d = 1 - R^d - T^d$. We use a generalized optical transfer-matrix approach [12] to calculate all relevant absorptances and obtain a good agreement of calculated and measured $A^d$, confirming the validity of the modeled $A_F^d$ [Fig. S1 (e)].

## Note 4. Error analysis.

The aim of the experiment is to resolve relatively small temporal shifts of emitted waveforms. Since the wedge design requires changing the position of the sample in order to vary $d$ it is crucial to analyze potential associated sources of experimental uncertainties. These are (i) the Gouy phase shift [13,14] of the emitted THz pulse, (ii) variations of the substrate thickness due to its lateral inhomogeneity, and (iii) long-term temporal drifts of the laser pulse train.

**Mitigation of uncertainties.** (i) We mitigate the potential Gouy phase shift, caused by a variation of $z$ position of the sample due to lateral scanning over the wedge, by setting the sample plane normal to the $z$-axis and the propagation direction of the excitation laser beam with an accuracy of $< 0.1°$. A lateral displacement of the sample over 3 mm results in an apparent temporal shift of $S(t)$ of less than 0.1 fs. (ii) The impact of a thickness variation of the substrate (ii) is ruled out by repeating same lateral scanning by 3 mm over the reference part of the sample where the Cu interlayer thickness $d = 0$, resulting in negligibly small apparent temporal shifts of <1 fs [see Fig. 3(a)]. This test also confirms no impact of Gouy phase shifts. Finally, (iii) the long-term stability of the laser pulse train was addressed by the measurement protocol: each waveform recorded from the wedge part of the sample ($d > 0$) was complemented by a subsequent measurement of a waveform emitted from the reference part of the sample ($d = 0$), serving as a temporal referencing. The stability of these reference waveforms was, however, excellent and of the order of 1 fs.

## Note 5. Spintronic origin of EO signals

The THz emission signals can be, in principle, contaminated by non-magnetic sources like any type of optical rectification or formation of bursts of electrical currents, possibly at the interfaces. In addition, even the magnetic signals do not have to originate from the spin currents and their conversion via ISHE, but from a magnetic dipole radiation. Below, we demonstrate are of magnetic and electric-dipole origin.

**Magnetic signals.** To remove any non-magnetic contributions, the THz emission signals are always measured for two opposite polarities of external magnetic field which flips the magnetization $\boldsymbol{M}$ of the ferromagnetic $F$ layer in our $\mathrm{F}|\mathrm{Cu}(d)|\mathrm{N}$ stack. We note that we are confident that $\boldsymbol{M}$ flips from our recent works with very similar magnetic layers, for example as shown in Ref. [18] of the main text. Typical raw emission signals $S^d_{+M}(t)$ and $S^d_{-M}(t)$ for positive and negative polarities of $\boldsymbol{M}$ and $d = 0$ $and$ 2.2 nm are shown in Fig. S3(a and c). Moreover, Fig. S3 (b and d) shows the difference signal, $S^-(d,t) = \left(S^d_{+M}(t) - S^d_{-M}(t)\right)/2$, and the sum signal, $S^+(d,t) = \left(S^d_{+M}(t) + S^d_{-M}(t)\right)/2$. The signal $S^-(d,t)$ captures all effects odd in $\boldsymbol{M}$, such as the conversion of spin currents by ISHE, and $S^+(d,t)$ all effect even in $\boldsymbol{M}$. As one can see, $S^-$ is larger by a factor of more than 200 and dominates the raw signals. However, we always plot the magnetic signals $S^-$ in the manuscript.

**Magnetic vs. electric dipole.** In principle, it needs to be shown that the magnetic signals $S^-$ are due to electronic spin currents (electric dipole) and not due to the ultrafast demagnetization of $\boldsymbol{M}$ (magnetic dipole), see Ref. [15] of the main text. The magnetic dipole can be ruled out by an optically symmetrized experiment where two samples are grown with opposite layer order and cap them by a window from the same material as the substrate. Upon the sample flip, the signals $S^-$ originating from the electric dipole (spin currents) is expected to reverse its polarity (it is odd in the layer order) while the components due to the magnetic dipole do not change the sign (are even in the layer order) [21] of the main text. As we see in Figs. 2 and 3 in the referenced paper, the emission from bilayers very similar to our reference sample flip perfectly its sign upon sample reversal. Alternatively, the same emission experiment can be performed in $F$ layer only. From these experiments, it can be concluded that the signals due to the magnetic dipole are generally significantly smaller than those due to spin currents, roughly by two orders of magnitude. Moreover, the magnetic dipole emission, normalized on the impedance $Z$ and absorptance $A$, cannot depend on the copper thickness. Therefore, we can summarize that the observed signals are due to electric dipole, i.e., due to electric spin-polarized currents.

## Note 6. Dynamical diffusion

**Dynamical diffusion.** In the absence of an external force and under the relaxation-time approximation, the occupation function of electrons is described by the Boltzmann transport equation [15,16]

$$\frac{\partial n_k(z,t)}{\partial t} + v_z \frac{\partial n_k(z,t)}{\partial z} = -\frac{(n_k - n_0)}{\tau} \tag{S5}$$

where $v_z = v_F \cos(\theta)$ is the projection of electron velocity into the $z$-direction, $n_0$ is the equilibrium Fermi-Dirac function, $n_k = n_0 + \Delta n_k$ is the non-equilibrium distribution and $\tau$ is the electron scattering time and $k$ is the wavevector. The electron-current density is given by

$$j = -e \int \frac{d^3 \boldsymbol{k}}{(2\pi)^3} v_z n_k, \tag{S6}$$

Using the diffusion approximations $\Delta n_k \ll n_0$ and $\frac{\partial}{\partial z} \Delta n_k \ll n_0$, we can insert Eq. (6) into Eq. (7) in the Fourier domain, which results in

$$j = -\frac{D_0}{1 - i\omega\tau} \frac{\partial N}{\partial z}, \tag{S7}$$

where $N$ is the electron density. This is the time-dependent dynamical Fick's law where the static diffusion coefficient $D_0 = \frac{v_F^2 \tau}{3}$ is replaced by the frequency-dependent diffusion coefficient $D(\omega) = \frac{D_0}{1 - i\omega\tau}$. Notably, $D(\omega)$ has the same form of the Drude conductivity.

**Wave-diffusion spin transport.** Here, we derive phenomenologically the full dispersion relation given in the Model section and Eq. (5) of the main text. However, a rigorous derivation is given in refs. [16,17].

We combine dynamical Fick's law in Eq. (S7) and the continuity equation for the spin transport $\partial_z j_s \propto -\partial_t \mu_s - 2\mu_s/\tau_{sf}$ [11,12] in the Fourier domain, yielding

$$j_s(z,\omega) = -D(\omega)\partial_z \mu_s(z,\omega) \tag{S8}$$

$$-i\omega \mu_s(z,\omega) + \partial_z j_s(z,\omega) + \frac{2\mu_s}{\tau_{sf}} = 0 \tag{S9}$$

where $\mu_s$ is the spin voltage as introduced in the main text, and $\tau_{sf}$ is the spin-flip time constant. Combining Eq. (S8) and (S9), one finds the dispersion relation

$$\bar{c}^2 k^2 = \omega^2 + i\omega \left(\frac{1}{\tau} + \frac{2}{\tau_{sf}}\right) - \frac{2}{\tau \tau_{sf}}, \tag{S10}$$

where $\bar{c}^2 = v_F^2/3$ is the mean propagation velocity and $k$ the wavevector. The spin current is given according to

$$j_s(z,t) = \int dt' \, \mu_s(t') P(z, t-t'), \tag{S11}$$

where the propagator is

$$P(z,t) = \int d\omega \, e^{i(\omega t - kz)}. \tag{S12}$$

The $P(z,t)$ captures the spin current propagation in Cu layer and $\mu_s(t)$ is the initial condition, i.e., the spin voltage generated in the ferromagnet after the optical pump excitation.

**References.**